\documentclass{iopconfser}
\usepackage{amsmath,amsfonts,amssymb}
\usepackage{epsfig}
\usepackage{times}
\usepackage{bm}
\usepackage[small,hang]{caption}
\usepackage[sort]{natbib}
\usepackage{subcaption}
\usepackage[percent]{overpic}
\usepackage{tikz}
\usetikzlibrary{arrows.meta, positioning}
\usepackage{ulem} 
\usepackage[T1]{fontenc}
\usepackage{lmodern}

\newcommand{\rgm}[1]{{\textcolor{black}{#1}}}
\newcommand{\jc}[1]{{\textcolor{black}{#1}}}
\newcommand{\rstrike}[1]{}
\newcommand{\jcstrike}[1]{}
\newcommand{\mylab}[3]{\raisebox{#2}[0mm][0mm]{\makebox[0mm][l]{\hspace*{#1}#3}}}

\begin{document}

\title{Interscale energy transfer in turbulent channels}

\author{Joy Chen$^{1}$ and Ricardo Garc{\'i}a-Mayoral$^{1}$}

\affil{$^1$Department of Engineering, University of Cambridge, Cambridge, CB2 1PZ, United Kingdom}

\email{r.gmayoral@eng.cam.ac.uk}

\begin{abstract}
We investigate the energy cascade in wall-bounded turbulence by analysing the interscale transfer between streamwise and spanwise length scales in 
periodic channels. This transfer originates from the nonlinear interactions in the advective term of the Navier–Stokes equations, which satisfy the 
classical triadic compatibility relations. Each triadic interaction is examined individually, and its corresponding nonlinear momentum and energy 
transfer are mapped to assess its relative importance in sustaining turbulence. Motivated by the anisotropy of the flow, we interpret each 
contribution $\partial_i(u_i u_j)$ to the advection term as carrying distinct physical information\rgm{, and therefore analyse them} separately.
Time-averaged maps of the energy transfer across all length scales and wall-normal positions for a channel flow at $Re_\tau \approx 180$ are used to
explore the \rstrike{possible} mechanisms underlying the cascade process. As a proof of concept, reduced-order simulations are performed by retaining only the
interactions \rgm{identified as} responsible for significant energy transfer based on our framework. Turbulent dynamics are successfully reproduced
when 30\% or more of the total interactions are included, while noticeable deviations emerge in the near-wall region when this proportion is further reduced.
\end{abstract}

\section{Introduction} 

\begin{figure}
    \centering
    \includegraphics[scale=0.25]{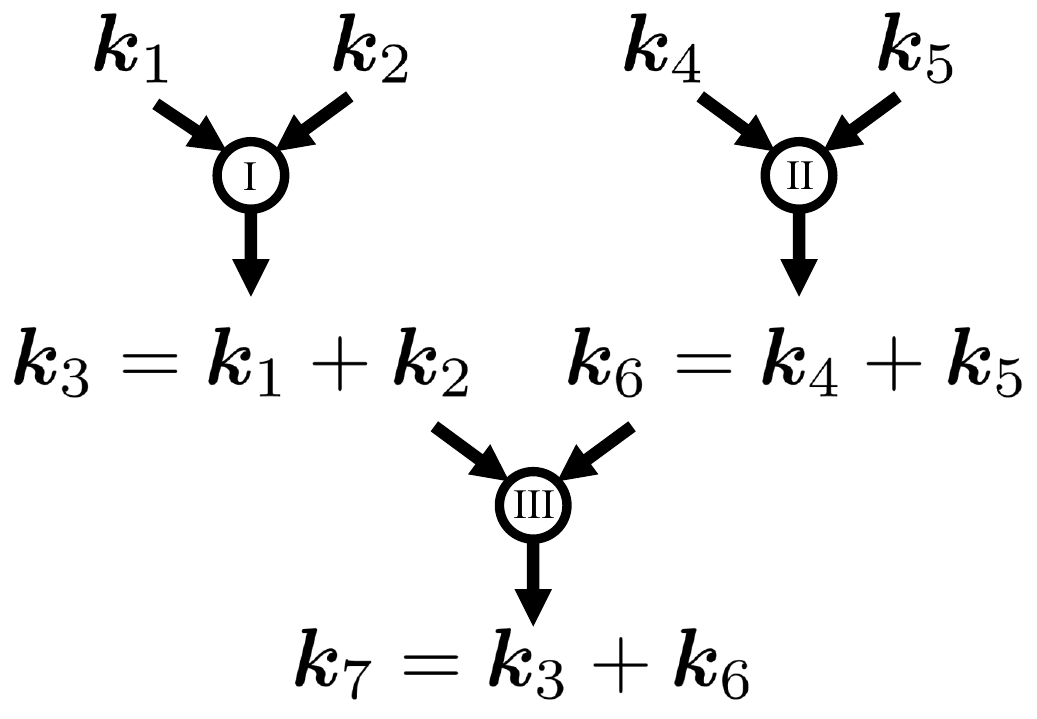}
    \caption{Sketch of interscale interactions\rgm{, adapted from \citet{de_salis_young_inter-scale_2024}}. Interaction \rgm{I} involves a recipient mode $\bm{k}_{3}$ receiving energy from an interaction between mode $\bm{k}_1$ and $\bm{k}_2$; $\bm{k}_{3}$ \rgm{in turn interacts} with $\bm{k}_{6}$ in interaction \rgm{III} to transfer energy into $\bm{k}_{7}$, and \rgm{so on to viscosity}.}
    \label{fig:hierachy_sketch}
\end{figure}

The turbulent energy cascade is central to our understanding of turbulence. In turbulent flows, energy is typically injected at large lengthscales and transferred through inter-scale interactions to smaller scales, ultimately being dissipated at viscous scales. In periodic channels, we can leverage the spatial periodicity of the flow to consider the governing equations in their streamwise- ($x$) and spanwise- ($z$) Fourier form. The inter-scale transfer is then inherently in the non-linear terms, as linear operators, though responsible for critical turbulent processes \citep{Lozano2021}, preserve wavenumber. In the Navier-Stokes equations, the non-linear term responsible for this transfer is the advection term, $\nabla \cdot (\bm{uu})$, which enables the triadic wavenumber interactions: a catalyst velocity component with wavenumber $\bm{k}_c$ advects a donor velocity component with wavenumber $\bm{k}_d$ to transfer momentum and energy to a recipient mode with wavenumber $\bm{k}_r = \bm{k}_c + \bm{k}_d$. The cascade proceeds through a hierarchy of such triadic interactions, as illustrated in Figure~\ref{fig:hierachy_sketch}, where different lengthscales can participate in the overall transfer with all three roles simultaneously, i.e. wavelength \rgm{$k_3$} serves as the recipient for interaction \rgm{I but} is also the catalyst for interaction \rgm{III}, etc. \rgm{The proposed framework emphasises the two-into-one directionality in each triadic interaction, i.e. catalyst and donor combining to cause an effect in a recipient, but makes} no \textit{a priori} distinction between interactions contributing to the direct cascade, energy transfer from large to small scales, and those contributing to the inverse cascade or backscatter.

The evolution of turbulent kinetic energy through inter-scale triadic interactions has been the subject of extensive research \citep{domaradzki_local_1990, domaradzki_nonlocal_1992, waleffe_nature_1992}. Previous studies have mapped triadic interactions in wall-bounded turbulence in various different forms. These include: maps of the bispectrum and biphase of streamwise wave triads \citep{cui_biphase_2021}; spanwise triadic wave interactions involved in turbulent energy transfer \citep{cho_scale_2018}; spanwise and streamwise triadic wave interactions involved in turbulent energy transfer \citep{lee_spectral_2019}; spanwise and streamwise wave dyad interactions into the principal resolvent mode \citep{bae_nonlinear_2021}; causal maps of inter-scale flux of information between signals of different wavelength \citep{Lozano2019};
\rgm{interscale triadic causality maps \citep{de_salis_young_inter-scale_2024}};
or the overall transfer from a donor mode to a recipient mode \citep{Ding2025}. Inter-scale transfer has also been analysed in the spatial domain through the use of structure functions and the Kármán–Howarth–Monin–Hill equation \citep{chiarini_ascendingdescending_2022, yao_analysis_2022}, and in the context of resolvent analysis when trying to characterise the non-linear, forcing term \citep{Zare2017,Morra2020,Symon2021}. \cite{Kawata2021} considered the interscale energy transfer in minimal channels, with streak breakdown and vortex regeneration viewed as instances of forward and reverse energy cascade.

\citet{de_salis_young_inter-scale_2024} drew a distinction between the nine individual contributions to the advective term in the Navier-Stokes equations, $\partial_i(u_i u_j)$. They argue\rgm{d} that, due to the anisotropy of wall turbulence, each of these contributions \rgm{could} be interpreted as representing a distinct physical mechanism.
They proposed a framework to map separately each of these contributions for each triadic-compatible set of modes\rstrike{, and verified a region of significant energy transfer against the known streak meandering behaviour of the near-wall cycle}.
We extend their preliminary exploration by generating maps that represent flow dynamics throughout the channel. We identify the key interactions responsible for significant energy transfer and, as a proof of concept, perform reduced-order simulations in which only these interactions are retained in the nonlinear advective term. This allows us to assess whether turbulent flow dynamics can be reproduced, thereby verifying the significance of the selected modes. As a benchmark, we use a channel of size $L_x=2\pi$, $L_z=\pi$,  $L_y=2$ at low friction Reynolds number, Re$_\tau\approx180$, leaving the consideration of larger and more complex flows for future work. 

The ultimate objectives of the proposed framework are threefold. First, to gain insights into the flow physics through the identification of the most significant nonlinear interactions. Second, to \rgm{provide} a \rgm{roadmap} for developing reduced-order models in which only a small number of key interactions are retained, thus reducing the computational cost of simulations \rgm{and proposing new approach to turbulence modelling}. Third, to identify potential ‘bottleneck’ regions in the interscale transfer maps that may play a critical role in the energy cascade. If such regions exist, they could be strategically targeted for optimal disruption of turbulence in flow\rgm{-}control applications.

The paper is organised as follows. \S 2 describes the methods used to map the inter-scale transfer in \rgm{an $x$-$z$-periodic} channel and the details of the reduced order simulations. \S 3.1 examines the maps of the inter-scale energy transfer to two different scales, and \S 3.2 compares the from the reduced order simulation to those of full DNSs. \S 4 summarises and concludes the paper. 

\section{Methods}
\subsection{Nonlinear energy transfer maps}
The Navier-Stokes and continuity equations in Fourier space for any given recipient mode $\bm{k}_r$ can be obtained by applying a streamwise ($x$)–spanwise ($z$) Fourier transform to the  equations in physical space,
\begin{equation}\label{eq:momentum}
\begin{split}
    \frac{\partial \bm{\hat{u}}}{\partial t}(\bm{k}_r, y, t) + 
    \!\!\!\!\!\sum_{\bm{k}_c + \bm{k}_d = \bm{k}_r} 
    \hat{\nabla} \cdot (\bm{\hat{u}}(\bm{k}_c, y, t)  \bm{\hat{u}}(\bm{k}_d, y, t)) = -\hat{\nabla} \hat{p}(\bm{k}_r, y, t) + \nu \hat{\nabla}^2 \bm{\hat{u}} (\bm{k}_r, y, t),
\end{split}
\end{equation}
\begin{equation}\label{eq:mass}
    \hat{\nabla} \cdot \bm{\hat{u}}(\bm{k}_r, y, t) = 0,
\end{equation}
where $\bm{u} = [u, v, w]^T$ is the velocity vector, $\nu$ is the kinematic viscosity and $p$ kinematic pressure. Fourier-transformed quantities are denoted with $\hat{(\cdot)}$ and $\hat{\nabla} = [\mathrm{i}k_x \; {\partial}/{\partial y} \; \mathrm{i}k_z]^T$ \rgm{is the Fourier transform of the del operator}. The linear operators ${\partial}/{\partial t}$, $\nabla$ and $\nabla^2$, along with the linear part of the advection $\hat{\nabla} \cdot (\bm{\hat{u}}(\bm{k}_r)\bm{\hat{U}})+\hat{\nabla} \cdot (\bm{\hat{U}}\bm{\hat{u}}(\bm{k}_r))$\rgm{,} where $\bm{U}$ represents the mean flow ($\bm{k}_d$ or $\bm{k}_c=0$), are Fourier transformed \rgm{trivially}. However, the non-linear part of the advection ($\bm{\hat{n}}$) becomes a convolution in the wavenumber domain, coupling velocity modes through triadic interactions, where

\begin{equation}\label{eq:LPN_deinitions}
\begin{split}
    \bm{\hat{n}}(\bm{k}_r) & = \sum_{\substack{\bm{k}_c + \bm{k}_d = \bm{k}_r \\ \bm{k}_c, \bm{k}_d \neq 0}} -\hat{\nabla} (\bm{\hat{u}}(\bm{k}_c) \bm{\hat{u}}(\bm{k}_d)).
\end{split}
\end{equation}

Each non-linear term for the recipient mode $\bm{k}_r$ is made up of all possible combinations of $\bm{k}_c$ and $\bm{k}_d$ such that $\bm{k}_c + \bm{k}_d = \bm{k}_r$. For each triadic combination of wavenumbers, the nonlinear term has nine contributions corresponding to the pairs of donor and catalyst velocity components,

\begin{equation}\label{eq:n_definition}
    \bm{\hat{n}}(\bm{k}_c, \bm{k}_d) = 
    \begin{bmatrix}
    \hat{n}_{uu} + \hat{n}_{vu} + \hat{n}_{wu} \\
    \hat{n}_{uv} + \hat{n}_{vv} + \hat{n}_{wv} \\
    \hat{n}_{uw} + \hat{n}_{vw} + \hat{n}_{ww}
    \end{bmatrix}= 
    \begin{bmatrix}
        -\mathrm{i} k_{t,x} \hat{u}(\bm{k}_c) \hat{u}(\bm{k}_d) - \frac{\partial}{\partial y} (\hat{v}(\bm{k}_c)\hat{u}(\bm{k}_d))
        -\mathrm{i} k_{t,z} \hat{w}(\bm{k}_c) \hat{u}(\bm{k}_d) \\
        - \mathrm{i} k_{t,x} \hat{u}(\bm{k}_c) \hat{v}(\bm{k}_d) 
        -\frac{\partial}{\partial y}(\hat{v}(\bm{k}_c) \hat{v}(\bm{k}_d))
        -\mathrm{i} k_{t,z} \hat{w}(\bm{k}_c) \hat{v}(\bm{k}_d) \\
        -\mathrm{i} k_{t,x} \hat{u}(\bm{k}_c) \hat{w}(\bm{k}_d) 
        -\frac{\partial}{\partial y}(\hat{v}(\bm{k}_c) \hat{w}(\bm{k}_d))
        -\mathrm{i} k_{t,z} \hat{w}(\bm{k}_c) \hat{w}(\bm{k}_d)
    \end{bmatrix}. 
\end{equation}

\citet{de_salis_young_inter-scale_2024} considered different metrics to quantify the nonlinear transfer. They showed that the mean-squared value\rstrike{$\langle|\hat{n}|^2\rangle$, where angle brackets denote time average,} is unable to account for the misalignment and cancellation between the phasors of $\hat{n}$ from different interactions into the same recipient \rgm{velocity}. Thus, they considered \rstrike{the} time-averaged momentum and energy transfer \rgm{metrics, making} use of the time-averaged projection $\mathcal{M}$ of \rgm{phasor} $b$ onto \rgm{phasor} $a$,
\begin{equation}
    \mathcal{M}(a, b) = \Re \langle a^* b \rangle,
\end{equation}
where $\Re$ denotes the real part. We can thus define the time-averaged change in momentum of the recipient wave,
\begin{equation}
        \left\langle  \frac{\partial }{\partial t} |\hat{u}(\bm{k}_r)| \right\rangle 
        = \mathcal{M}\left(e_u(\bm{k}_r), \frac{\partial \hat{u}(\bm{k}_r)}{\partial t}\right),
    \label{eq:momentummetric}
\end{equation}
where $e_u=\hat{u}(\bm{k}_r)/|\hat{u}(\bm{k}_r)|$\rgm{,} and its time-averaged change in energy,
\begin{equation}
        \left\langle  \frac{\partial }{\partial t} \left(\frac{1}{2}|\hat{u}(\bm{k}_r)|^2 \right)\right\rangle 
        = \mathcal{M}\left(\hat{u}(\bm{k}_r), \frac{\partial \hat{u}(\bm{k}_r)}{\partial t}\right).
    \label{eq:energymetric}
\end{equation}
The latter gives rise to the standard energy metric stemming from energy budget analysis, while the former is intrinsically connected to the momentum equation~\ref{eq:momentum}. For a fixed recipient mode, the two metrics produce the same results with only a \rgm{scaling} factor $|\hat{u}(\bm{k}_r)|$. However, when comparing across different recipient modes, only the energy metric provides a meaningful scale for comparison. As such, we use the energy metric for the present analysis. The nonlinear energy transfer due to a triad where velocity component $u_i$ advects $u_j$ can be computed with $\mathcal{M}(\hat{u}_j(\bm{k}_r), \hat{n}_{ij}(\bm{k}_c, \bm{k}_d))$ using flow fields from existing DNS datasets.

\begin{figure}
    \begin{center}
    \includegraphics[width=0.64\linewidth]{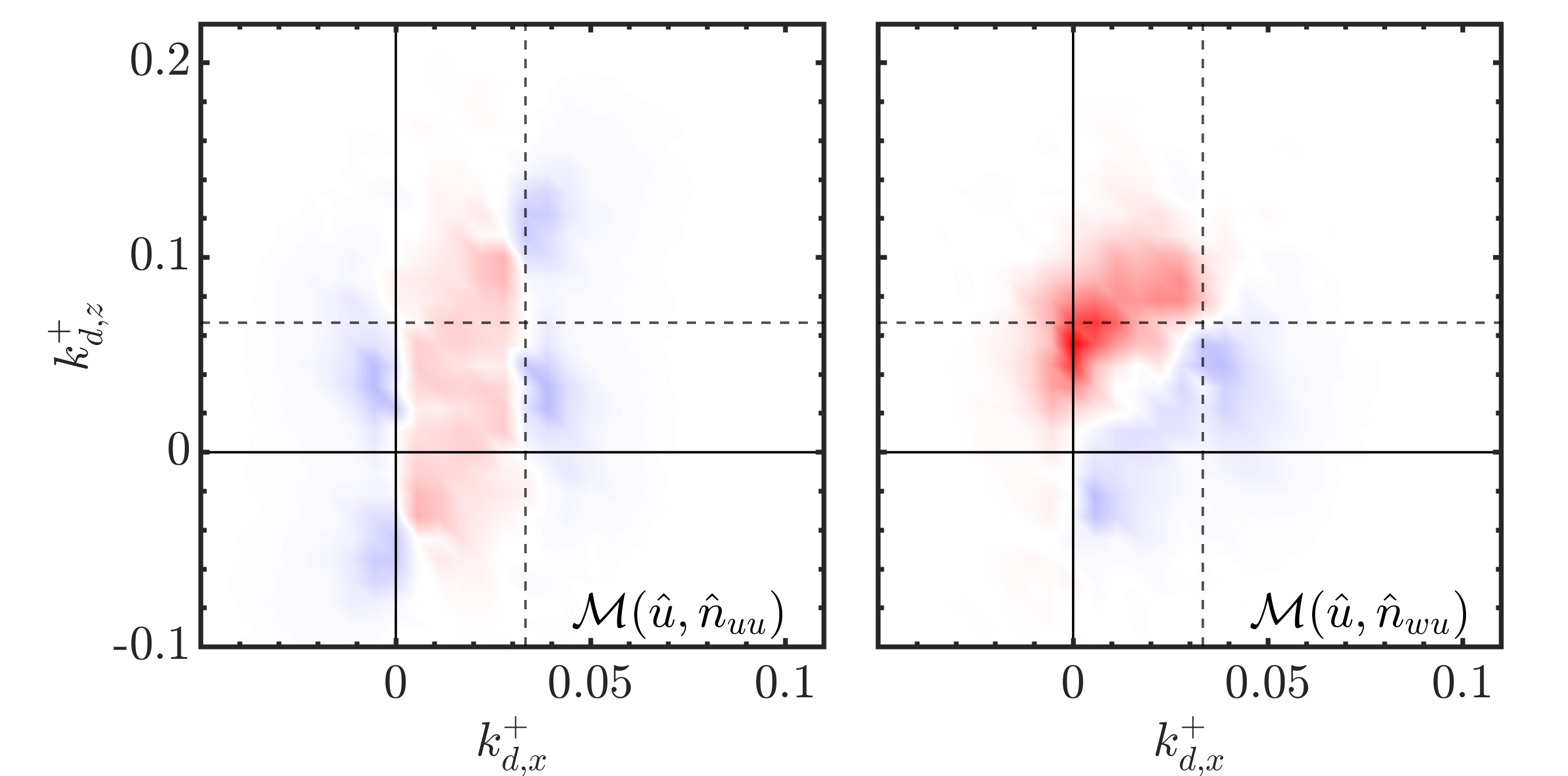}
    \mylab{-105mm}{46mm}{(a)}
    
    \vspace*{3mm}
    
    \includegraphics[width=0.64\linewidth]{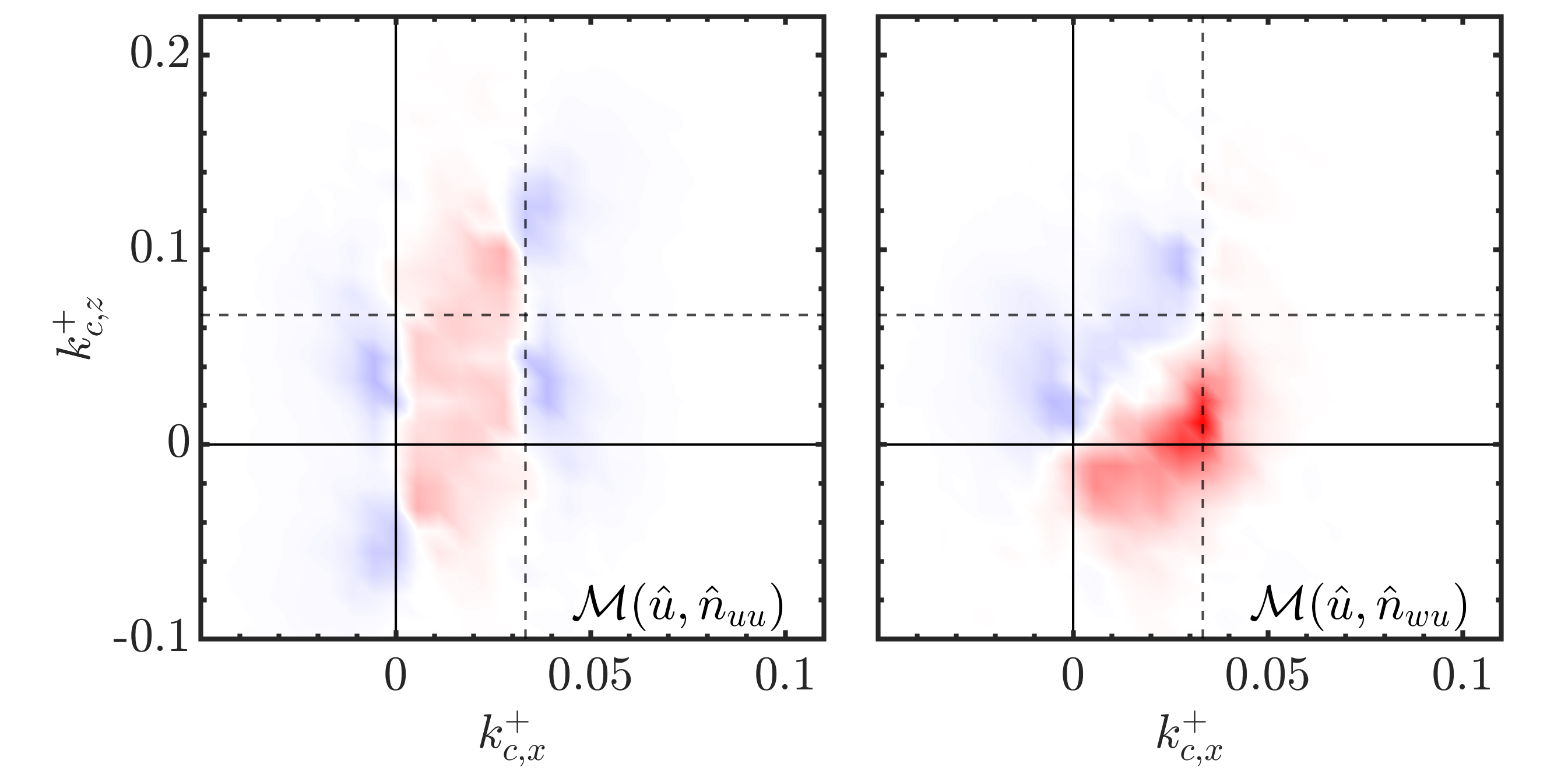}
    \mylab{-105mm}{47mm}{(b)}

    \caption{Maps of inter-scale energy transfer in (a) donor and (b) catalyst wavenumbers to recipient lengthscale  $\lambda_{t,x}^+ = 188, \lambda_{t,z}^+ = 94$ at $y^+ = 15$ from the $uu$ and $wu$ contributions to the advection term for an $Re_\tau = 180$ channel flow. The colour scale in wall units is from -0.1 (blue) to 0.1 (red). The black dashed lines indicate the recipient wavenumber $\bm{k}_r$. }
    \label{fig:kp_2_7x7}
    \end{center}
\vspace{7mm}
    \centering
    \begin{overpic}[scale=0.7]{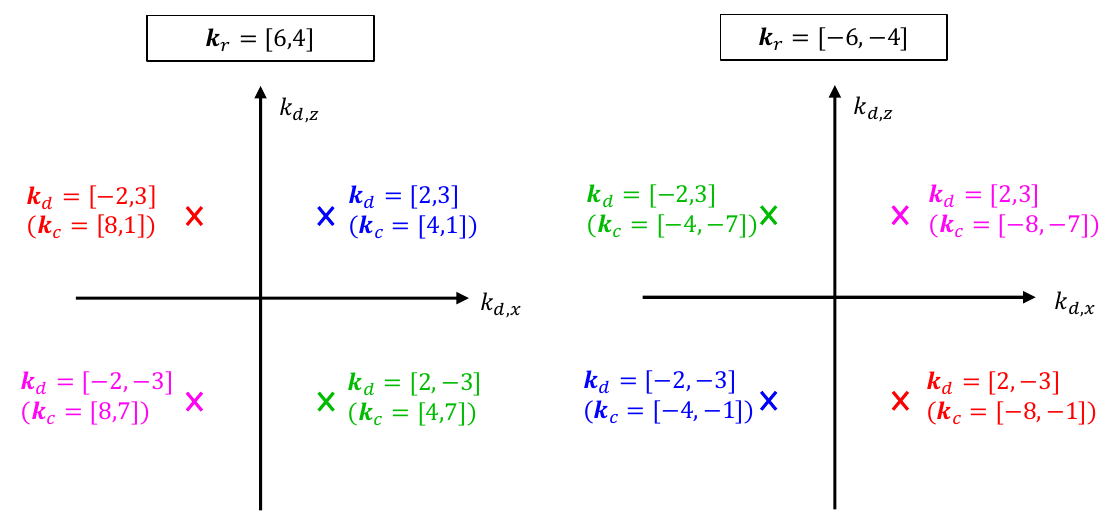}
        \put(0,45){(a)} 
        \put(50,45){(b)}   
    \end{overpic}
    \caption{Illustration of triadic interactions represented in each quadrant of the maps for two recipient modes with opposite signs. Modes of the same colour represent interactions between the same lengthscales, with the only difference being the sign of the wavenumber. }
    \label{fig:folding}
\end{figure}

For a given recipient wavenumber of interest, \citet{de_salis_young_inter-scale_2024} generated 2D maps of inter-scale transfer which can be portrayed in terms of donor wavenumber $\bm{k}_d = [k_{d,x}, k_{d,z}]^T$, with the catalyst wavenumber given implicitly by $\bm{k}_c = \bm{k}_r - \bm{k}_d$, or vice versa. In wavenumber space, there is no practical difference between expressing the maps in terms of $\bm{k}_c$ or $\bm{k}_d$\rgm{, as} the $\bm{k}_c$ map is simply the $\bm{k}_d$ map flipped across both axes and shifted by $\bm{k}_r$. An example of these maps for two of the nine components, $\hat{n}_{uu}$ and $\hat{n}_{wu}$, presented in terms of catalyst and donor, is shown in Figure~\ref{fig:kp_2_7x7}. 

\rgm{In contrast,} when representing the maps in log-wavelength space, as commonly done for spectra and cospectra, care must be taken in the treatment of positive and negative wavenumbers. \rgm{For the pair $\pm k$,} spectra typically combine their contributions into a single wavelength $\lambda = 2\pi/k$, which is appropriate since physical waves arise from the combination \rgm{of both wavenumbers}.
However, as shown in Figure~\ref{fig:kp_2_7x7}, the four quadrants of the interscale map are not symmetric, and such merging may lead to a loss of
\rgm{distinction in value and} even cancellation between asymmetric contributions, obscuring the significance of \rgm{contributions in different} quadrants.

\begin{figure}[t!]
    \centering
    \definecolor{darkgreen}{rgb}{0.0,0.9,0.0}
    \begin{overpic}[scale=0.9]{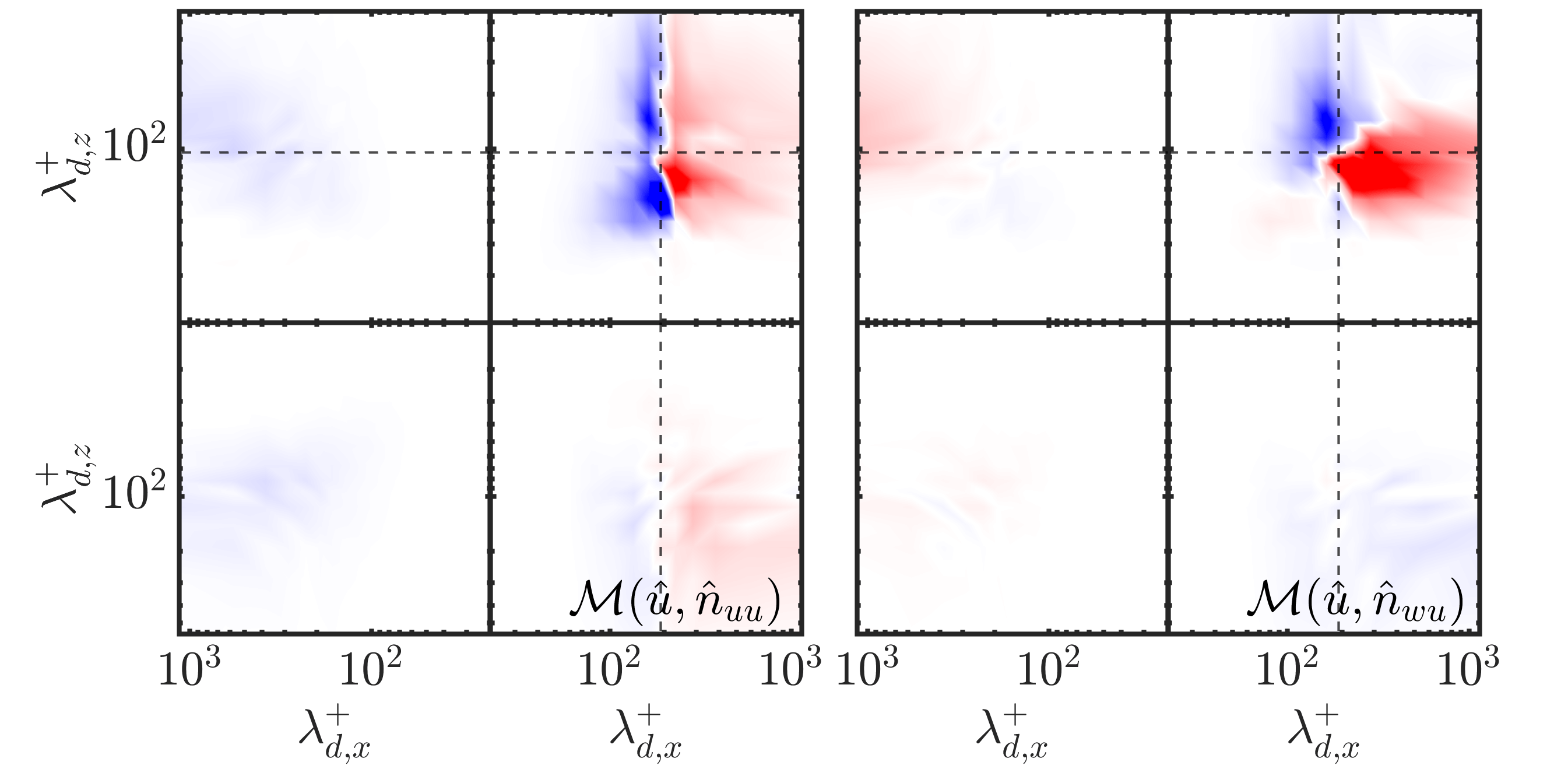}
            \put(2,48){(a)} 
            \put(42,36){\tikz{\draw[black,line width=1pt] (0,0) circle (4pt);}} 
            \put(86,37.5){\tikz{\draw[black,line width=1pt] (0,0) ellipse (10pt and 4pt);}} 
    \end{overpic}

    \vspace*{4mm}
    
    \begin{overpic}[scale=0.9]{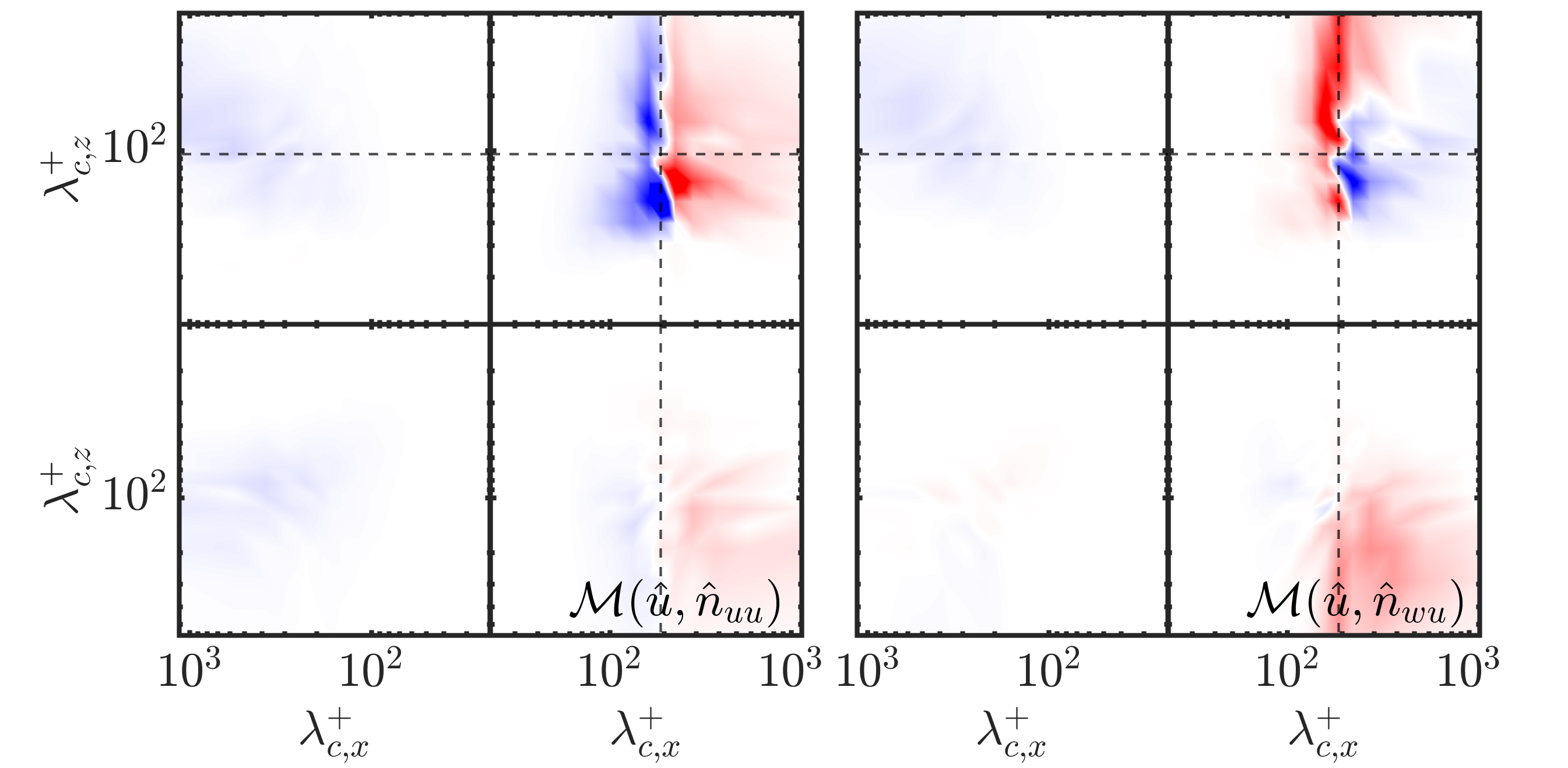}
            \put(2,48){(b)} 
            \put(46,10){\tikz{\draw[black,line width=1pt] (0,0) circle (6pt);}}
            \put(82.5,44){\tikz{\draw[black,line width=1pt] (0,0) ellipse (8pt and 6pt);}} 
    \end{overpic}
    \caption{Log-$\lambda$ maps of inter-scale energy transfer in (a) donor and (b) catalyst modes to recipient lengthscale  $\lambda_{t,x}^+ = 188, \lambda_{t,z}^+ = 94$ at $y^+ = 15$ from the $uu$ and $wu$ contributions to the advection term for an $Re_\tau = 180$ channel flow. The colour scale in wall units is from $-2\times10^{-5}$ (blue) to $2\times10^{-5}$ (red), scales based on methods detailed in \S \ref{sec:scaling}. The black dashed lines indicate the recipient wavenumber $\bm{k}_r$. Circled regions in black in the corresponding maps of (a) and (b) represent the same interactions.}
    \label{fig:logl_2quad_7x7}
\end{figure}
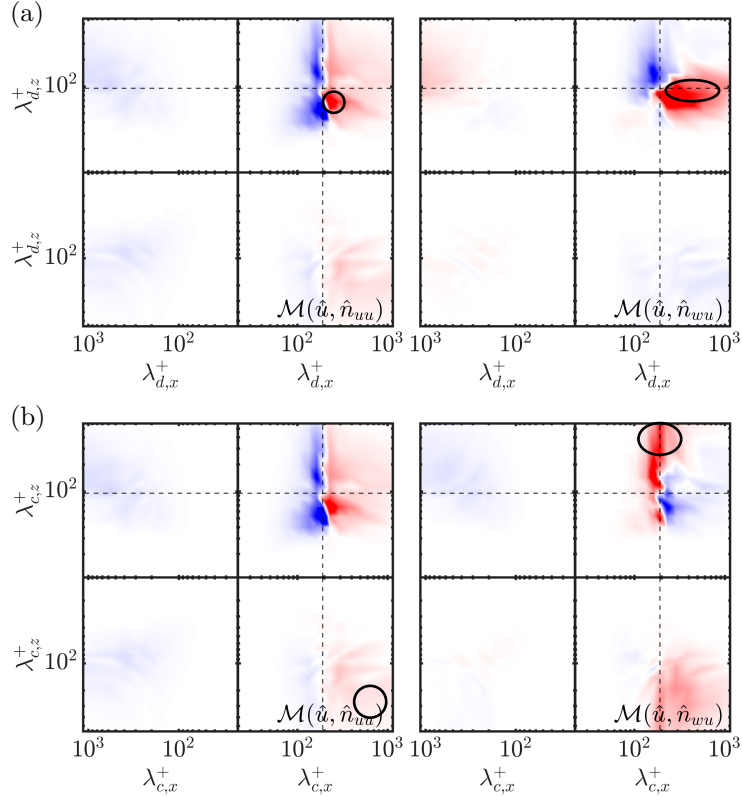

The asymmetry occurs because points on each of the quadrants of the map represent a distinct triadic interaction, \rgm{as} illustrated in figure~\ref{fig:folding}(a).  For example, the interaction of $\bm{k}_d = [2,3]$ into $\bm{k}_r = [6,4]$ requires a catalyst of $\bm{k}_c = [4,1]$, while if the donor mode is $\bm{k}_d = [-2,3]$, the catalyst \rgm{is} $\bm{k}_c = [8,1]$.
These two interactions, along with the corresponding \rgm{ones} in the two other quadrants, have donors and recipients of \rgm{equal wavelengths}, but \rgm{involve} catalysts \rgm{of different wavelengths}, thus corresponding to four \rgm{physically different mechanisms}.
Interactions between physical waves, comprising both positive and negative wavenumbers, which justifies the combination of the quadrants in a typical spectrum, arise when the sign change is applied consistently to both $\bm{k}_d$ and $\bm{k}_r$. Comparing between figures~\ref{fig:folding}(a) and (b), which have recipient modes of opposite sign, we notice interactions that include modes of the same \rgm{wavelengths} (coloured the same) but shown in different quadrants. These two maps \rgm{can} thus be combined after rotating one by $180^\circ$ without the loss of information, and this can similarly be done for maps of $\bm{k}_r = [-6,4]$ and $[6,-4]$. Thus, {only one map} is required for each recipient lengthscale, \rgm{but it must retain all four quadrants,} with each converted separately into log-wavelength space.
\rgm{In our maps, we omit} the negative signs \rgm{in wavelengths for simplicity}, and the axis directions are reversed to preserve the visual layout of four quadrants, as illustrated in figure\rstrike{s}~\ref{fig:logl_2quad_7x7}. 
\jc{For} \rgm{maps given in terms of catalyst wavelengths $\lambda_{c,i}$, like the ones in figure} \jc{\ref{fig:logl_2quad_7x7}(b)} \rgm{, the corresponding $\lambda_{d,i}$ is given by
\begin{equation}
    \lambda_{d,i} = \frac{1}{\left|1\mp\lambda_{r,i}/\lambda_{c,i}\right|}\lambda_{r,i},
\end{equation}
where $i$ refers to either $x$ or $z$ and the sign in the denominator depends on the quadrant for $\lambda_{c,i}$. In turn, for maps given in terms of donor wavelengths the corresponding $\lambda_{c,i}$ is given by
\begin{equation}
    \lambda_{c,i} = \frac{1}{\left|1\mp\lambda_{r,i}/\lambda_{d,i}\right|}\lambda_{r,i}.
\end{equation}
}

The construction of log-wavelength maps also involves premultiplication \rgm{by} $\bm{k}_c$ or $\bm{k}_d$, which makes maps expressed in terms of the catalyst or the donor wavelength provide different information.
Circled regions \rgm{in the figure} correspond to the same interactions\rgm{, and} illustrate \rgm{how the different premultiplications undergone can result in}
very different magnitudes. We \rgm{note} in particular \rgm{that} the maps \rgm{for the same velocity component in catalyst and donor,} $\partial(u_iu_i)/\partial x_i$\rgm{,} are visually the same\rgm{,} since $\hat{\nabla} (\hat{u}(\bm{k}_c)\hat{u}(\bm{k}_d)) = \hat{\nabla} (\hat{u}(\bm{k}_d) \hat{u}(\bm{k}_c))$.
However, the same points on the maps do not correspond to the same interaction; instead, there is a swap in the roles of donor and catalyst. \rgm{Altogether,} the maps in figure~\ref{fig:logl_2quad_7x7}(a) and (b) \rgm{can be considered} to contain complementary information.


\subsection{Reduction of data size with spectral binning}

\begin{figure}
\begin{center}
\includegraphics[width=0.7\linewidth]{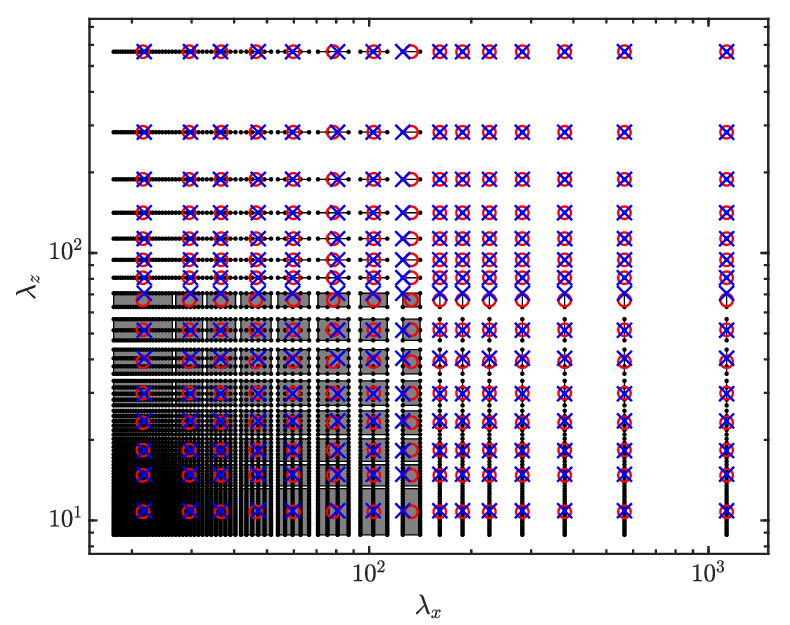}
    \caption{Spectral binning procedure illustrated in log–$\lambda$ space. Black points indicate the computed modes, and grey regions represent the selected bins. Red circles mark the geometric centres where the averaged data are stored, while the blue cross denotes the mode closest to each geometric centre. }
    \label{fig:spectralBinning}
\end{center}
\end{figure}

To identify the key interactions throughout the channel, maps are required for recipient wavenumbers spanning the full DNS-resolved range and extending across the entire channel height. However, this full set for a channel with $\mathrm{Re}_\tau \approx 180$ would produce a dataset roughly 20,000 times \rgm{the size of one instantaneous flow field} for \jc{one velocity component}. To make the analysis plausible, we apply the spectral binning procedure of \citet{jimenez_turbulent_2010}, in which the information for groups of high-wavenumber modes, forming the densely populated region at short wavelengths in a log–wavelength spectrum, is combined as illustrated in figure \ref{fig:spectralBinning}. The values within each bin are averaged and represented at the geometric centre of the interval, defined as $k = \sqrt{k_1 k_2}$ in one dimension, where $k_1$ and $k_2$ denote the start and end wavenumbers of the bin, respectively. 

Similar procedures are applied to reduce the number of recipient modes for which maps are constructed. Since all modes that act as catalyst or donor 
modes in a map may also serve as recipient modes, the black points in figure~\ref{fig:spectralBinning} also represent the full range of recipient 
modes considered. We select the recipient modes closest to the geometric centre of each bin for map construction and assume that the nonlinear
interactions into all recipient modes within a bin are comparable to that of the selected mode. In the wall-normal direction, rather than computing maps
across the entire DNS $y$ grid, a reduced subset of $y^+$ locations is chosen to generate representative maps that capture the flow dynamics in the key regions of the channel: $y^+ = 5,10,15,25,40,60,90,130$ and $180$.  

Overall, this procedure reduces the dataset size for the maps to roughly the equivalent of \rgm{10} flow fields, making the analysis far more tractable. This number scales approximately with $Re_\tau^2$, depending on the number of wavenumbers after binning, which does not scale linearly with total wavenumbers in each direction. For $Re_\tau \approx 550$, for example, the full dataset would be around 200,000 times the \rgm{size of one} flow field, and the reduced one based on the \rgm{above binning method would be around} 100 times \rgm{the size of one flow field}.

\subsection{Scaling across recipient \rgm{modes} and heights}\label{sec:scaling}
\begin{figure}[!t]
    \centering
    \begin{subfigure}{0.48\linewidth}
        \begin{overpic}[width=\linewidth]{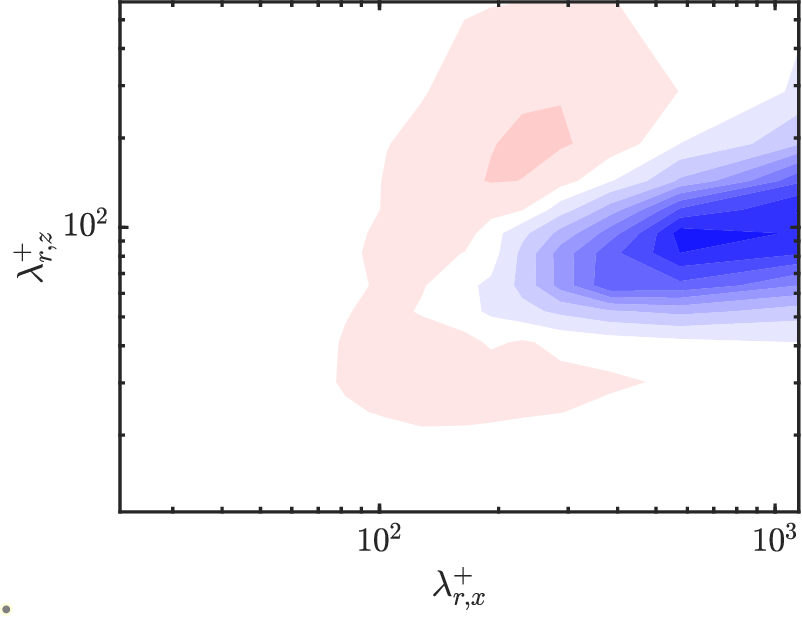}
            \put(3,70){(a)} 
        \end{overpic}
    \end{subfigure}
    \hspace{0.1cm}
    \begin{subfigure}{0.48\linewidth}
        \begin{overpic}[width=\linewidth]{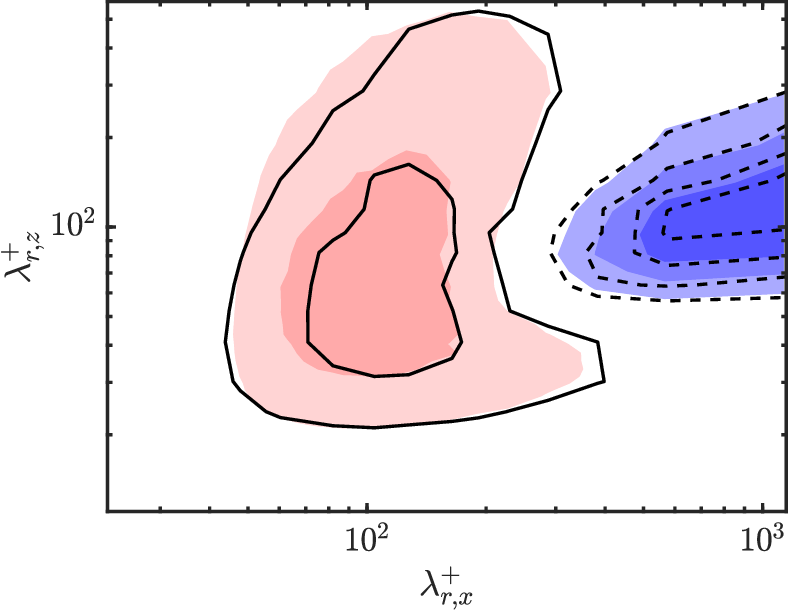}
            \put(3,70){(b)}
        \end{overpic}
    \end{subfigure}
    \vspace*{-3mm}
    \caption{\jc{(a) Total nonlinear energy transferred ($\hat{N}$) into each recipient \rgm{wavelength} at $y^+ = 15$. Coloured contours from blue to red are from -0.05 to 0.05, with increments of 0.005, in viscous units. (b) Total nonlinear energy transferred into each recipient \rgm{wavelength} integrated \rgm{in $y$} across the channel. \rgm{Solid and dashed contours are results for the present channel, and 
    coloured ones} data from \citet{Ding2025}. Contour levels are from -1.4 to 1.4 with increments of 0.2.}}
    \label{fig:scaling}
\end{figure}

To enable meaningful comparison between maps corresponding to different recipient wavenumbers at a given wall-normal plane, an appropriate \rgm{overarching} scaling is required.
\rgm{The energy metric of equation \ref{eq:momentummetric} enables integration} over all nine contributions ($\partial_i(u_i u_j)$) of each map \rgm{to yield} the total nonlinear energy transferred into the corresponding recipient wavenumber
\rgm{-- note that a similar sumation property is not available using the momentum metric of equation \ref{eq:energymetric}}.
Figure \ref{fig:scaling}(a) shows, for $y^+ = 15$, the energy transferred into each recipient lengthscale, where each data point represents the integral of the maps for one given recipient wavenumber. Since this spectral map is itself presented in log–wavelength space, each data point is premultiplied by its associated recipient wavenumber. This premultiplication provides the necessary scaling for comparing results across \rgm{maps for different target wavelengths}.
Overall, each map \rgm{then} requires two stages of premultiplication: first by the catalyst or donor wavenumbers \rgm{of the map axes,} to \rgm{portray the map in} log–wavelength space, and second by the recipient wavenumber\rgm{,} for consistent scaling \rgm{across the whole set of recipients. We note
that the latter does not alter the appearance of the maps, e.g. those portrayed in figure~\ref{fig:logl_2quad_7x7}, but allows for quantitative
comparison of maps for different recipient wavelengths.}
\rgm{We also note that the procedure is equivalent to that used in \cite{Ding2025} to produce maps of energy transfer into each recipient wavelength integrated in $y$ across the whole channel. Our results for e.g. figure \ref{fig:scaling}(a) collapse with theirs once integrated in $y$, as shown in figure \ref{fig:scaling}(b), validating the present mapping procedure.}


\begin{figure}[!t]
    \centering
    \vspace*{2mm}
    \includegraphics[width=0.5\linewidth]{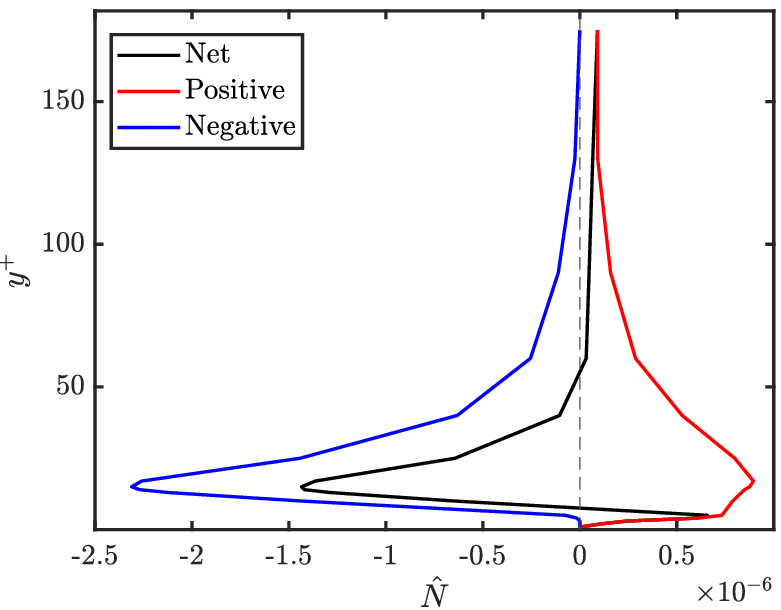}
    \vspace*{-1mm}
    \caption{\jc{Total nonlinear energy transferred at each $y^+$. Red and blue lines are the total positive and negative energy transferred from integrating only the red or blue parts of \ref{fig:scaling}(a).}}
    \label{fig:yscale}
\end{figure}

Integrating \rgm{maps like that in} figure \ref{fig:scaling}(a) \rgm{across all recipient wavelengths} yields the total nonlinear energy transfer at \rgm{any given height $y$}, and performing the same procedure across \rgm{the entire $y$ range} produces figure \ref{fig:yscale}. The intensity of these net nonlinear interactions decreases markedly with distance from the wall. The relative importance of advection and nonlinear transfer, however, increases far away from the wall, where viscous effects are negligible. Consequently, $y$-scaling is necessary to ensure that the significance of interscale transfer in the channel core is accurately represented.
\rgm{The total nonlinear energy transfer at each height
cannot be used directly for this scaling, as it reverses sign and more importantly becomes zero at some heights. We therefore use}
the total positive energy transfer, obtained by summing only the positive contributions to the energy transferred into the recipient modes, such as the region in red in figure \ref{fig:scaling}(a).

\subsection{Reduced order simulations}\label{sec:method-reducedorder}

 We \rgm{next} conduct reduced-order simulations in which only the nonlinear interactions identified as important from our maps are retained. \rgm{If turbulence is sustained and remains essentially the same as in a full DNS, this would suggest} that the selected interactions are indeed physically meaningful.

We conduct reduced order simulations at $\mathrm{Re}_\tau\approx180$ by applying a constant mean pressure gradient. The channel is of size $2\pi\delta \times \pi \delta \times \delta$ in the streamwise, spanwise and wall-normal direction. The numerical code is adapted from \citet{fairhall_spectral_2018},
in which the three-dimensional incompressible Navier–Stokes equations are solved using a spectral discretisation in the streamwise and spanwise directions, while the wall-normal direction is discretised using second-order finite differences on a staggered grid. The wall normal grid is stretched with $\Delta y_{min}^+ \approx 0.3$ near the wall and $\Delta y_{max}^+ \approx 3$ in the centre of the channel. The wall-parallel resolutions are $\Delta x^+ \approx 6$ and $\Delta z^+ \approx 3$. The temporal integration uses a Runge–Kutta discretisation, where pressure is corrected to reinforce incompressibility. Every time step is divided into three substeps, each of which uses a semi-implicit scheme for the viscous terms and an explicit scheme for the advective terms. Once the flow reaches a statistically steady state, statistics are collected for at least $10\delta/u_\tau$. 

In a standard DNS simulation, the nonlinear term is computed by transforming the velocity fields to physical space via Fourier transforms, multiplying the components, and transforming back. To manage each triadic interaction individually, we instead compute the nonlinear term in Fourier space using a convolution, which allows calculation of only a selected set of wavenumber interactions from the maps obtained above. 

The set of interactions is chosen by applying a \jcstrike{$y$-dependent}threshold on the nonlinear energy transferred, such that only interactions with energy transfer above this threshold are retained. Since maps based on catalyst and donor {wavelengths} contain \rgm{different} information, all interactions 
\rgm{above the threshold in either map} are included. By varying the selection threshold, we \rgm{can vary the percentage of interactions 
retained} relative to a full DNS. In this study, simulations were conducted with $15\%, 30\%, 45\%$ and $60\%$ of all interactions. 

The premultiplication of the modes \rgm{with $k_x=0$ or $k_z=0$}, i.e., the modes with infinite wavelength in either the streamwise or spanwise direction, requires careful consideration.
Due to the limited size of our domain, $L_x = 2\pi$ and $L_z = \pi$, energy in wavelengths larger than the domain is transferred to \rgm{the corresponding $k_x=0$ or $k_z=0$ modes} \citep{lozano-duran_effect_2014}.
In a larger channel, where these modes were resolved, \rgm{the interscale maps would include their wavelengths and the corresponding premultiplication factors would be straightforward.}
However, in the present domain, all interactions involving the zero modes are, in principle, premultiplied by zero, effectively rendering their contribution negligible. Some of these interactions, however, may be critical for sustaining turbulence.

We leave for future work to perform this analysis in larger domains, where all relevant modes are sufficiently resolved \rgm{and all relevant interactions have non-zero premultiplication.} 
As a preliminary approach, we instead premultiply these modes by $k_1/2$, where $k_1$ is the first non-zero mode, rather than by 0.
\rgm{Our reasoning is that, considering a channel size twice as large as the present one, some of the energy that is in $k = 0$ in the present channel
would become resolved into wavenumber $k = k_1/2$, while some remained in $k = 0$. Premultiplying all this energy by $k_1/2$ could then result in us
keeping unimportant interactions, but not in us neglecting some important ones by accident. 
We note, however, that a more conservative approach would entail premultiplying by $k_1$} \jc{as,} \rgm{for channels of any size larger than the present one, $k_1$ would be the overall upper bound for the wavenumber of modes with wavelengths larger than $\lambda_1$.}

\section{Results}
\subsection{Nonlinear energy transfer maps}

\begin{figure}[!t]
    \centering
    \hspace{-3mm}
    \includegraphics[scale=1]{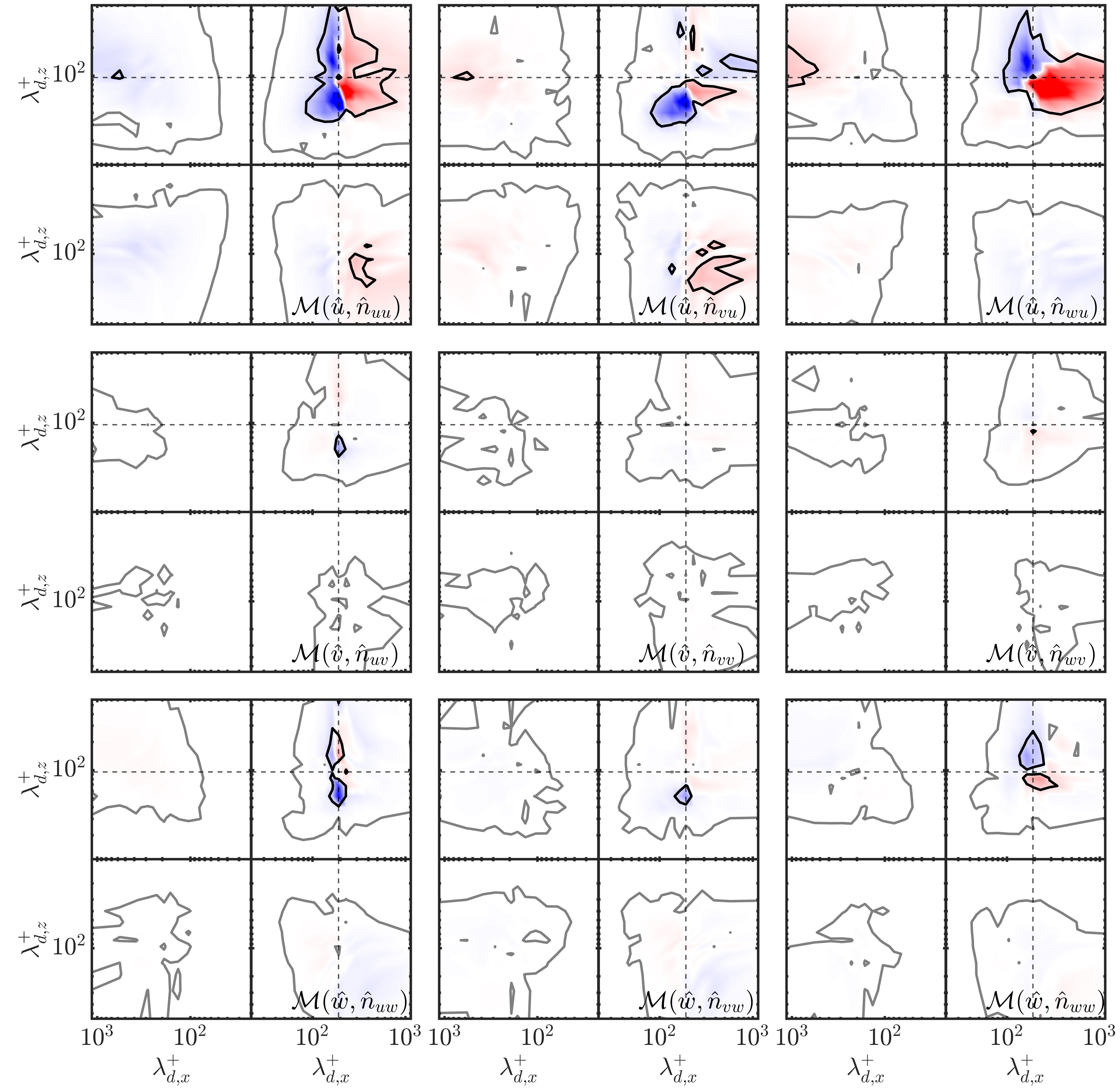}
    \caption{Maps of inter-scale energy transfer in donor and log-wavelength space to recipient lengthscale  $\lambda_{t,x}^+ = 188, \lambda_{t,z}^+ = 94$ at $y^+ = 15$ from the nine contributions to the advection term for an $Re_\tau = 180$ channel flow. The colour scale in wall units is from $-5\times10^{-5}$ (blue) to $5\times10^{-5}$ (red). The black dashed lines indicate the recipient wavenumber $\bm{k}_r$. Black and grey contour lines represent interactions selected if 0.5\% and 15\% of all interactions are kept, respectively. }
    \label{fig:quad_7x7_p}
\end{figure}

\begin{figure}[!t]
    \centering
    \hspace{-3mm}
    \includegraphics[scale=1]{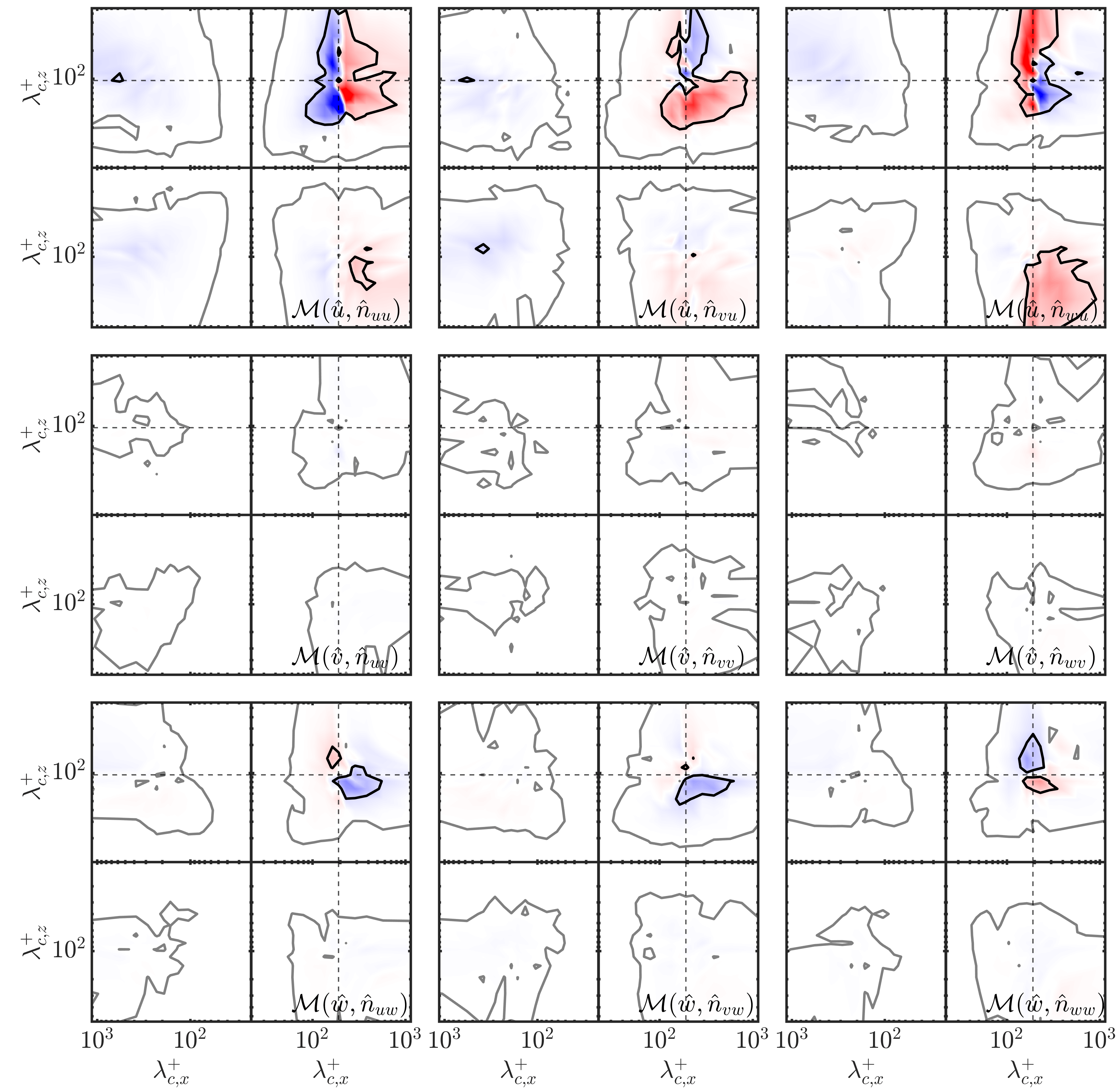}
    \caption{Maps of inter-scale energy transfer in catalyst log-wavelength space, parameters of the graph are the same as those in Figure~\ref{fig:quad_7x7_p}. }
    \label{fig:quad_7x7_a}
\end{figure}

\citet{de_salis_young_inter-scale_2024} investigated the significant interactions into the recipient lengthscale of $\lambda_{r,x}^+ = 188, \lambda_{r,z}^+ = 94$ at $y^+ = 15$ as a representative scale of the near-wall cycle, using maps in wavenumber space such as those in figure~\ref{fig:kp_2_7x7}. Their results suggested that the dominating mechanism of interscale transfer was the streak meandering, where spanwise-velocity ($w$) structures with a streamwise length similar to the size of the recipient mode but much longer in span advected streamwise-velocity ($u$) structures with a spanwise length similar to the size of the recipient mode but much longer in $x$, i.e. the streaks, resulting in a transfer of energy into \rgm{$u$ in the} recipient mode and causing the \rgm{initially-}$x$-elongated streaks to become sinuous.
This interaction also appears significant in \rgm{the present premultiplied maps in log-wavelength space, i.e. in the panel for $\mathcal{M}(\hat{u},\hat{n}_{wu})$} in figures~\ref{fig:quad_7x7_p} and \ref{fig:quad_7x7_a}.
However, \rgm{the new} maps highlight additional interactions that are similarly significant, such as the intense regions in \rgm{the panels for} $\mathcal{M}(\hat{u},\hat{n}_{uu})$ and $\mathcal{M}(\hat{u},\hat{n}_{vu})$. 

Across both donor and catalyst maps, the most significant interactions are concentrated in the first quadrant, roughly centred around the recipient \rgm{wavelength} and, in some cases, extending towards infinitely large \rgm{ones}.
In terms of wavenumbers, these dominant regions lie approximately at $\bm{k} = \bm{k}_r \pm \bm{k}_r$. Consequently, the third mode in the triad spans
from $\bm{k} = 0$ to $\bm{k} = \jc{\pm}\bm{k}_r$ ($\bm{\lambda}_r \lesssim \bm{\lambda} \lesssim \infty$), which explains the moderate energy levels observed in these regions.
This indicates that interactions between scales comparable to and larger than the recipient, regardless of which acts as donor \rgm{and which as} catalyst, are generally the most significant contributors to energy transfer. Such behaviour is consistently observed across different recipient modes. This trend aligns with the \rgm{general depiction of the energy cascade as going from larger to smaller scales}, though the presence of substantial negative energy transfer suggests a non-negligible backscatter from the \rgm{recipients to larger scales}.

\rgm{The black solid contour in the maps in figures~\ref{fig:quad_7x7_p} and \ref{fig:quad_7x7_a}} encloses the regions of \rgm{most intense energy transfer}. When this threshold is applied consistently across all maps at different wall-normal positions and recipient scales \rgm{as detailed in \S~\ref{sec:scaling}}, the enclosed region \rgm{contains roughly} 0.5\% of all interactions. \rgm{We would expect that retaining these modes alone} would be sufficient to sustain turbulence, but as shown later, a substantially larger fraction is required.
\rgm{For illustration, the larger region delimited} by the grey contour line contains roughly 15\% of all interactions\rgm{, which is not yet sufficient to sustain healthy turbulence}.
\jc{Note that the above 0.5\% and 15\% apply to the interactions across the whole channel; the percentage in any specific plane can vary.
Note also that the percentage \rgm{number} of modal interactions retained is not necessarily the same as the percentage area enclosed in the map, as \rgm{in logarithmic maps modes are more widely spaced for higher wavelengths.}}



\subsection{Reduced order simulations}

\begin{figure}[!t]
\begin{center}
\includegraphics[width=\linewidth]{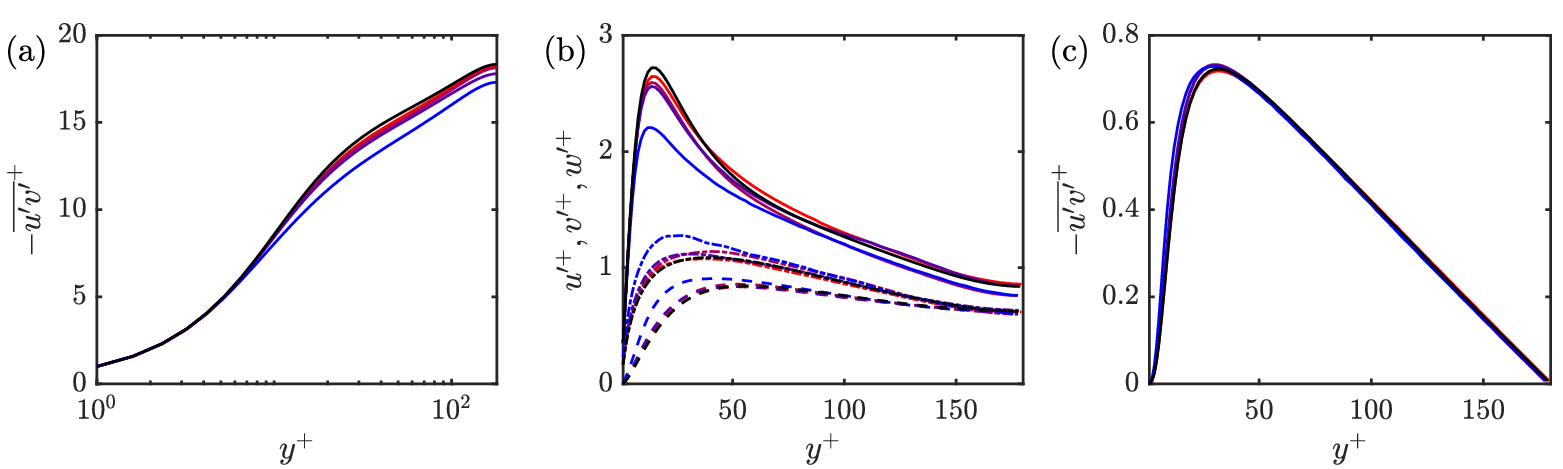}
\vspace*{-6mm}
    \caption{Comparison of (a) mean flow profile, (b) rms of velocity fluctuation and (c) Reynolds stress of reduced order
    simulations.  Black: full DNS, blue to red: reduced order simulations computing 15\% to 60\% of all interactions.}
    \label{fig:1dstats}
\end{center}
\vspace{2mm}
\centering
\begin{overpic}[width=\linewidth]{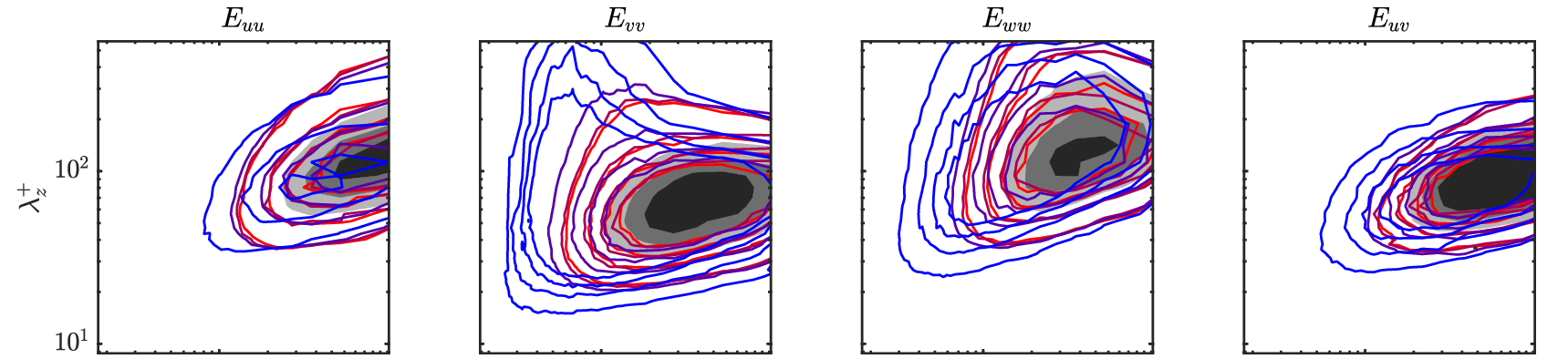}
    \put(2,20){(a)}
\end{overpic}
\begin{overpic}[width=\linewidth]{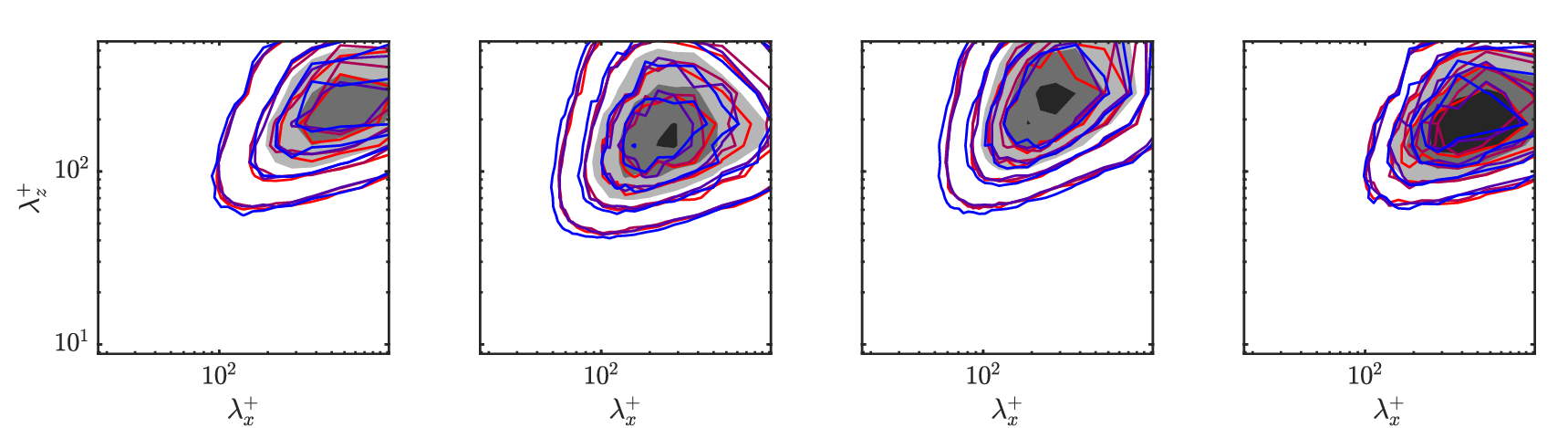}
    \put(2,25){(b)}
\end{overpic}
\vspace*{-6mm}
    \caption{Comparison of the spectral energy densities of the three velocities and Reynolds stress at (a) $y^+=15$ and (b) $y^+ = 100$. Filled contours are from full DNS, contour line colours are as in figure \ref{fig:1dstats}. Contour levels are the same across the different cases. }
    \label{fig:spectrum}
\end{figure}

\begin{figure}
\begin{center}
\includegraphics[width=\linewidth]{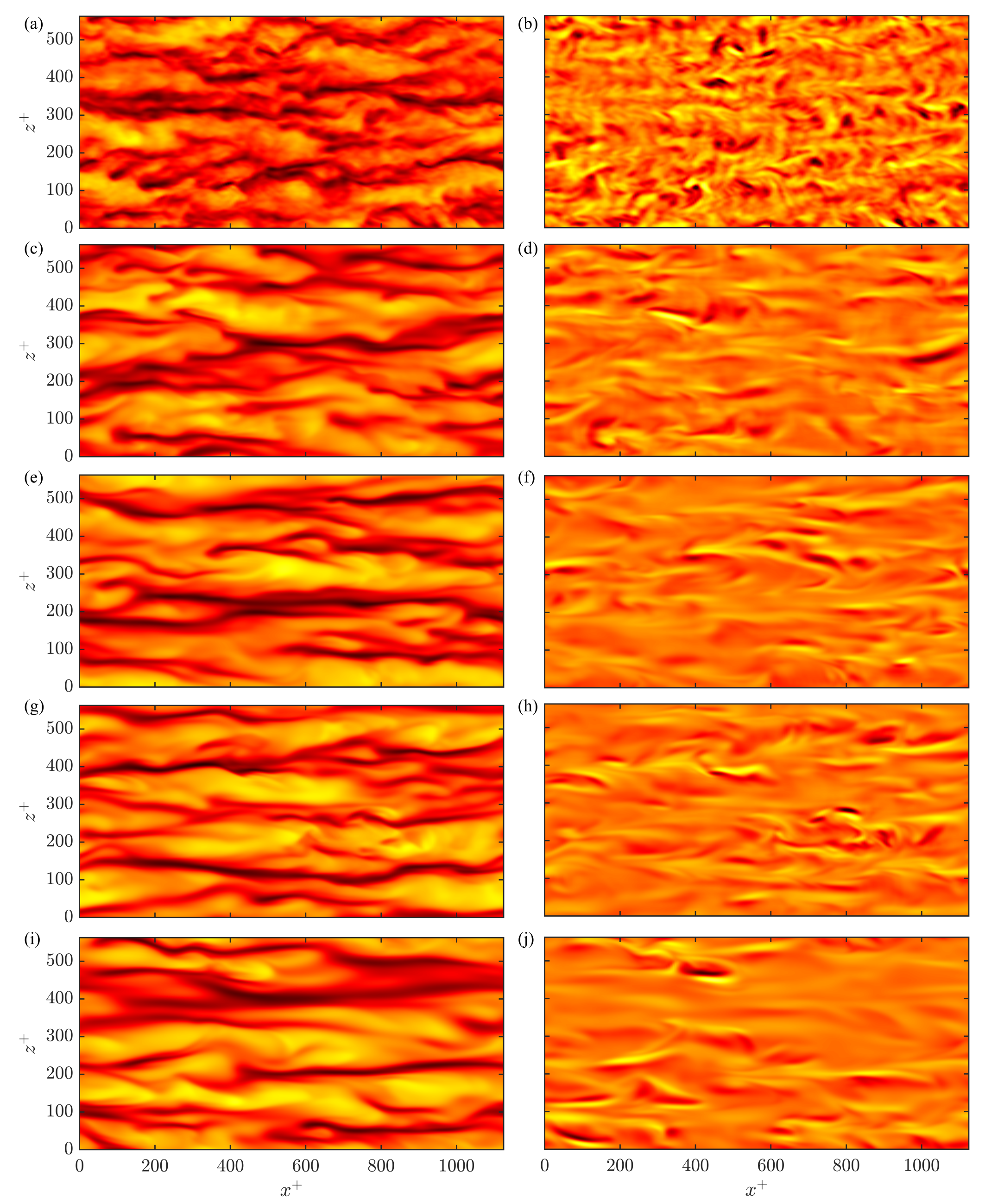}
    \caption{\rgm{
    Instantaneous realisation of streamwise $u^+$ (left column) and wall-normal $v^+$ (right column) velocity at
    $y^+ = 15$. (a-h), reduced order simulations, from top to bottom with $15\%, 30\%, 45\%$ and $60\%$ of all interactions;
    (i-j), standard DNS. Colours from dark to clear are from 4 to 20 for $u^+$ and from -3 to 3 for $v^+$.}}
    \label{fig:snapshots}
\end{center}
\end{figure}

\jc{Results for the reduced order simulations, retaining from 15\% to 60\% of interactions compared to regular DNS, are presented below. The simulations are conducted following the methods detailed in \S~\ref{sec:method-reducedorder}. }

The turbulent statistics from the reduced order simulations are compared with the full DNS simulation in figure~\ref{fig:1dstats}. The mean \rgm{velocity profile} and \rgm{the} Reynolds shear stress across all reduced-order simulations align well with the regular DNS results\rgm{, suggesting that this modelling approach may be of use to predict bulk flow properties}. For the velocity fluctuations, however, simulations retaining 30\% of the interactions or more show \rgm{good agreement with} the full DNS, but for 15\% significant deviations appear. Attempts to use a lower percentage, e.g. 5\%, resulted in the \rgm{simulations} diverging, and are thus not presented in the figure. 

Premultiplied energy spectra are portrayed at $y^+ = 15$ and $y^+ = 100$ in figure \ref{fig:spectrum}. \rgm{Other than} for the 15\% case at $y^+ = 15$,
\rgm{there is} a good \rgm{collapse} in the \rgm{regions of high spectral density} for all variables. \rgm{The reduced-order simulations} have \rgm{however} higher intensity than the full DNS, which \rgm{appears not to be consistent with the rms values portrayed} in figure~\ref{fig:1dstats}.
The \rgm{reason is that the} streamwise and spanwise infinite modes, which are \rgm{omitted} in the spectra, include significantly less energy than \rgm{in the full} DNS, \rgm{resulting overall in a smaller change} in the 1D statistics.
This \rgm{likely} occurs due to our treatment of interactions involving modes with $k_x$ or $k_z = 0$\jc{, as discussed in \S~\ref{sec:method-reducedorder}}\rgm{; further investigation into the strategy for selecting which of these modes to retain is left for future work.} 
\rgm{For the simulation retaining 15\% of the interactions, more} significant deviations are observed at $y^+ = 15$ across all \rgm{the} spectra. There is an overall shift toward shorter $\lambda_x$, and \rgm{for} $E_{vv}$ \rgm{also} a widening in the spanwise direction, a feature reminiscent of flows over textured surfaces such as riblets, rough or porous substrates \citep{rgm11,abderrahaman2019,GG19,Hao2025}.
At $y^+ = 100$, \rgm{in contrast, good} agreement \rgm{with the full DNS is observed across all reduced-order simulations\jcstrike{, both} in terms of characteristic lengthscales\jcstrike{ and intensity}.}
The excess of small-scale energy near the wall \rgm{may be associated with} excessive nonlinear energy transfer into the \rgm{overexcited} modes,
coupled with insufficient transfer \rgm{away from them}, resulting in \rgm{spurious} energy accumulation.
This \rgm{suggests that the omitted interactions would have a non-negligible net dissipative contribution into these modes,
and that omitting them results in an energy imbalance for these recipient wavelengths}.
We leave \rgm{for future investigation} the task of ensuring net nonlinear energy balance \rgm{for these} modes. 

\rgm{The agreement or disagreement between reduced-order simulations and full DNS observed at $y^+ = 15$ in the spectra can also be observed in instantaneous flow realisations, as shown in figure~\ref{fig:snapshots}.
The flow fields of simulations retaining 30\% or more of the interactions are visually similar to those of the full DNS.
For the simulation retaining 15\%, however, the typical streamwise-elongated signatures in $u$ and $v$ are disrupted and shortened, particularly for $v$, for which they also become somewhat spanwise elongated at small $\lambda_x$, in agreement with the observation on the $E_{vv}$ spectrum in figure \ref{fig:spectrum}(a).
}

\rgm{Overall, the simulations with reduced interactions}
successfully reproduce the key dynamics of turbulent flow. However, even at 15\%, a substantial number of interactions that appear weak or negligible in the maps of figures~\ref{fig:quad_7x7_p} and \ref{fig:quad_7x7_a} are still included. This suggests that the current \rgm{thresholding method} is far from \rgm{yielding} the \rgm{truly} minimal set of essential interactions. \rgm{The present work supports the hypothesis that only a fraction of all the possible interscale interactions need to be retained to sustain healthy turbulence, but} further \rgm{investigation is} needed to systematically reduce the percentage of interactions retained\rgm{, and} to better assess the  efficiency \rgm{and usefulness} of the proposed framework \rgm{for modelling purposes}.


\section{Conclusions}
In this work, we have proposed a framework for the analysis of interscale interactions in wall turbulence by focusing on how information is transferred between $x$- and $z$-lengthscales in the Navier-Stokes momentum equations. Inter-scale transfer is possible only through the non-linear advection terms, in which velocity fluctuations of lengthscale $\bm{k}_c$ advect velocities of lengthscale $\bm{k}_d$ to cause a transfer of energy to a recipient lengthscale $\bm{k}_r$. There are three catalyst and donor velocity components, which correspond to nine contributions to the advection term. Given the high anisotropy of wall turbulence, the proposed framework discriminates between these nine contributions as representing different physical mechanisms even within the same triadic set. We propose to use this framework to identify the dominant advection mechanisms through which energy is transferred between different lengthscales.

\rgm{To identify and quantify the intensity of the dominant interactions governing the turbulent energy cascade, we have generated maps for the} time-averaged interscale energy transfer across the entire channel.
The validity of this framework was assessed through reduced-order simulations, in which the nonlinear term was computed in Fourier space via convolutions but restricted to only the selected interactions.
For cases retaining 30\% or more of the total \rgm{number of} interactions, the resulting turbulent statistics match \rstrike{reasonably}well those from full DNS.
\rstrike{However,}Noticeable deviations appear near the wall when only 15\% of the interactions are retained, and further reduction leads to numerical instability.
The 15\% \rgm{simulation exhibited} an accumulation of energy \rgm{in} short streamwise wavelengths, suggesting that the reduced set of interactions causes a substantial alteration in the nonlinear energy transfer into those modes. 

While the good representation of turbulence with 30\% of the interactions is a promising result, many apparently weak interactions are still included, which suggests that efforts are needed for further minimising the retained interactions and isolating the essential mechanisms of energy transfer in wall-bounded turbulence.

\section*{Acknowledgements}
This work was supported in part by the European Research Council under the Caust grant ERC-AdG-101018287. Joy Chen acknowledges the support of the Bassil Fund from Selwyn College, University of Cambridge.
\rgm{We are thankful to Jitong Ding, Daniel Chung and Simon Illingworth for kindly providing their data portrayed in figure \ref{fig:scaling}.}

\bibliography{causality}

@article{abderrahaman2019,
  title={Modulation of near-wall turbulence in the transitionally rough regime},
  author={Abderrahaman-Elena, N. and Fairhall, C. T. and Garc{\'\i}a-Mayoral, R.},
  journal={J. Fluid Mech.},
  volume={865},
  pages={1042--1071},
  year={2019}
}

@article{bae_nonlinear_2021,
	title = {Nonlinear mechanism of the self-sustaining process in the buffer and logarithmic layer of wall-bounded flows},
	volume = {914},
	pages = {A3},
	journal = {Journal of Fluid Mechanics},
	author = {Bae, H. Jane and Lozano-Durán, A. and McKeon, Beverley J.},
	year = {2021},
}

@article{chiarini_ascendingdescending_2022,
	title = {Ascending–descending and direct–inverse cascades of {Reynolds} stresses in turbulent {Couette} flow},
	volume = {930},
	journal = {Journal of Fluid Mechanics},
	author = {Chiarini, Alessandro and Mauriello, Mariadebora and Gatti, Davide and Quadrio, Maurizio},
	year = {2022},
	pages = {A9},
}

@article{cho_scale_2018,
	title = {Scale interactions and spectral energy transfer in turbulent channel flow},
	volume = {854},
	journal = {Journal of Fluid Mechanics},
	author = {Cho, Minjeong and Hwang, Yongyun and Choi, Haecheon},
	year = {2018},
	pages = {474--504},
}

@article{cui_biphase_2021,
	title = {Biphase as a diagnostic for scale interactions in wall-bounded turbulence},
	volume = {6},
	number = {1},
	journal = {Physical Review Fluids},
	author = {Cui, G. and Jacobi, I.},
	year = {2021},
	pages = {014604},
}

@article{Ding2025,
  title = {Mode-to-mode nonlinear energy transfer in turbulent channel flows},
  volume = {1002},
  journal = {Journal of Fluid Mechanics},
  author = {Ding,  Jitong and Chung,  Daniel and Illingworth,  Simon J.},
  year = {2025},
  pages = {A42}
}

@article{domaradzki_local_1990,
	title = {Local energy transfer and nonlocal interactions in homogeneous, isotropic turbulence},
	volume = {2},
	issn = {0899-8213},
	number = {3},
	journal = {Physics of Fluids A: Fluid Dynamics},
	author = {Domaradzki, J. Andrzej and Rogallo, Robert S.},
	year = {1990},
	pages = {413--426},
}

@article{domaradzki_nonlocal_1992,
	title = {Nonlocal triad interactions and the dissipation range of isotropic turbulence},
	volume = {4},
	issn = {0899-8213},
	number = {9},
	journal = {Physics of Fluids A: Fluid Dynamics},
	author = {Domaradzki, J. Andrzej},
	year = {1992},
	pages = {2037--2045},
}

@article{rgm11,
  title={Hydrodynamic stability and breakdown of the viscous regime over riblets},
  author={Garc{\'i}a-Mayoral, Ricardo and Jim{\'e}nez, Javier},
  journal={J. Fluid Mech.},
  volume={678},
  pages={317--347},
  year={2011}
}

@article{GG19,
  title={Turbulent drag reduction by anisotropic permeable substrates--analysis and direct numerical simulations},
  author={{G{\'o}mez-de-Segura}, Garazi and Garc{\'i}a-Mayoral, Ricardo},
  journal={J. Fluid Mech.},
  volume={875},
  pages={124--172},
  year={2019}
}

@article{Hao2025,
  title = {Turbulent flows over porous and rough substrates},
  volume = {1008},
  journal = {J. Fluid Mech.},
  author = {Hao,  Zengrong and García-Mayoral,  Ricardo},
  year = {2025},
  pages = {A1}
}

@article{Kawata2021,
  title = {Scale interactions in turbulent plane Couette flows in minimal domains},
  volume = {911},
  journal = {Journal of Fluid Mechanics},
  publisher = {Cambridge University Press (CUP)},
  author = {Kawata,  Takuya and Tsukahara,  Takahiro},
  year = {2021},
  pages = {A55}
}

@article{lee_spectral_2019,
	title = {Spectral analysis of the budget equation in turbulent channel flows at high {Reynolds} number},
	volume = {860},
	journal = {Journal of Fluid Mechanics},
	author = {Lee, Myoungkyu and Moser, Robert D.},
	year = {2019},
	pages = {886--938},
}

@article{Lozano2019,
  title = {Causality of energy-containing eddies in wall turbulence},
  volume = {882},
  journal = {Journal of Fluid Mechanics},
  author = {Lozano-Durán,  Adrián and Bae,  H. Jane and Encinar,  Miguel P.},
  year = {2019},
  pages = {A2}
}

@article{Lozano2021,
  title = {Cause-and-effect of linear mechanisms sustaining wall turbulence},
  volume = {914},
  journal = {Journal of Fluid Mechanics},
  author = {Lozano-Durán,  Adrián and Constantinou,  Navid C. and Nikolaidis,  Marios-Andreas and Karp,  Michael},
  year = {2021},
  pages = {A8}
}

@article{Morra2020,
  title = {The colour of forcing statistics in resolvent analyses of turbulent channel flows},
  volume = {907},
  journal = {Journal of Fluid Mechanics},
  author = {Morra,  Pierluigi and Nogueira,  Petr\^onio A. S. and Cavalieri,  André V. G. and Henningson,  Dan S.},
  year = {2020},
  pages = {A24}
}

@article{Symon2021,
  title = {Energy transfer in turbulent channel flows and implications for resolvent modelling},
  volume = {911},
  journal = {Journal of Fluid Mechanics},
  author = {Symon,  Sean and Illingworth,  Simon J. and Marusic,  Ivan},
  year = {2021},
  pages = {A3}
}

@article{waleffe_nature_1992,
	title = {The nature of triad interactions in homogeneous turbulence},
	volume = {4},
	number = {2},
	journal = {Physics of Fluids A: Fluid Dynamics},
	author = {Waleffe, Fabian},
	year = {1992},
	pages = {350--363},
}

@article{yao_analysis_2022,
	title = {Analysis of interscale energy transfer in a boundary layer undergoing bypass transition},
	volume = {941},
	journal = {Journal of Fluid Mechanics},
	author = {Yao, H. and Mollicone, J.-P. and Papadakis, G.},
	year = {2022},
	pages = {A14},
}

@article{Zare2017,
  title = {Colour of turbulence},
  volume = {812},
  journal = {Journal of Fluid Mechanics},
  author = {Zare,  Armin and Jovanović,  Mihailo R. and Georgiou,  Tryphon T.},
  year = {2017},
  pages = {636--680}
}

@inproceedings{de_salis_young_inter-scale_2024,
    title = {Inter-scale {Causality} {Relations} in {Wall} {Turbulence}},
    volume = {2753},
    doi = {10.1088/1742-6596/2753/1/012019},
    language = {en},
    urldate = {2025-05-10},
    booktitle = {5th {Madrid} {Turbulence} {Workshop}},
    author = {De Salis Young, J and Hao, Z and Garcia-Mayoral, R},
    month = apr,
    year = {2024},
    pages = {012019},
}

@article{jimenez_turbulent_2010,
    title = {Turbulent boundary layers and channels at moderate {Reynolds} numbers},
    volume = {657},
    copyright = {https://www.cambridge.org/core/terms},
    issn = {0022-1120, 1469-7645},
    doi = {10.1017/S0022112010001370},
    language = {en},
    urldate = {2025-10-14},
    journal = {Journal of Fluid Mechanics},
    author = {Jiménez, Javier and Hoyas, Sergio and Simens, Mark P. and Mizuno, Yoshinori},
    month = aug,
    year = {2010},
    pages = {335--360},
}

@article{lozano-duran_effect_2014,
    title = {Effect of the computational domain on direct simulations of turbulent channels up to {$Re_\tau$} = 4200},
    volume = {26},
    issn = {1070-6631},
    doi = {10.1063/1.4862918},
    number = {1},
    urldate = {2025-08-06},
    journal = {Physics of Fluids},
    author = {Lozano-Durán, Adrián and Jiménez, Javier},
    month = jan,
    year = {2014},
    pages = {011702},
}

@article{fairhall_spectral_2018,
    title = {Spectral {Analysis} of the {Slip}-{Length} {Model} for {Turbulence} over {Textured} {Superhydrophobic} {Surfaces}},
    volume = {100},
    issn = {1386-6184, 1573-1987},
    doi = {10.1007/s10494-018-9919-1},
    language = {en},
    number = {4},
    urldate = {2025-10-30},
    journal = {Flow, Turbulence and Combustion},
    author = {Fairhall, C. T. and García-Mayoral, R.},
    month = jun,
    year = {2018},
    pages = {961--978},
}

\end{document}